\input{aipcheck}

\documentclass[
    ,final            
  ]
  {aipproc}

\layoutstyle{6x9}

\newcommand{\newc}{\newcommand}
\newc{\gsim}{\lower.7ex\hbox{$\;\stackrel{\textstyle>}{\sim}\;$}}
\newc{\lsim}{\lower.7ex\hbox{$\;\stackrel{\textstyle<}{\sim}\;$}}

\def\beq{\begin{equation}}
\def\eeq{\end{equation}}
\def\beqn{\begin{eqnarray}}
\def\eeqn{\end{eqnarray}}

\def\half{{\textstyle{1\over 2}}}

\def\ie{{\it i.e.}\/}
\def\eg{{\it e.g.}\/}

\def\IR{\relax{\rm I\kern-.18em R}}
 \font\cmss=cmss10 \font\cmsss=cmss10 at 7pt
\def\IQ{\relax{\rm I\kern-.18em Q}}
\def\IZ{\relax\ifmmode\mathchoice
 {\hbox{\cmss Z\kern-.4em Z}}{\hbox{\cmss Z\kern-.4em Z}}
 {\lower.9pt\hbox{\cmsss Z\kern-.4em Z}}
 {\lower1.2pt\hbox{\cmsss Z\kern-.4em Z}}\else{\cmss Z\kern-.4em Z}\fi}

\begin{document}

\title{New Non-Trivial Vacuum Structures in Supersymmetric Field Theories
         \thanks{
          Talk given by B.T. at the XIII Mexican School of Particles and Fields, held
           in San Carlos, Mexico, October 2008.}}


\author{Keith R. Dienes, $\,$Brooks Thomas}{
  address={Department of Physics, University of Arizona, Tucson, AZ  ~85721~ USA}
}

\begin{abstract}
    In this talk, we present 
     three examples of new non-trivial vacuum structures
    that can occur 
    in supersymmetric field theories, along with explicit models in which they arise.
    The first vacuum structure is one in which
    supersymmetry is broken at tree-level in a perturbative
    theory that also contains a supersymmetry-preserving ground state.  
    Models realizing this structure are uniquely characterized by the fact that
    no flat directions appear in
    the classical potential, all vacua appear at finite distances 
    in field space, and no non-perturbative physics is required for vacuum stability.
    The second non-trivial vacuum structure we discuss consists of large
     (and even infinite) towers of metastable vacua, and we show that 
    models which give rise to 
    such vacuum towers exhibit a rich set of 
    instanton-induced vacuum tunneling dynamics.
    Finally, our third new non-trivial vacuum structure consists of
    an infinite number of degenerate vacua;  this leads to
     a Bloch-wave ground state and a vacuum ``band'' structure.
    Models with such characteristics therefore experience time-dependent vacuum oscillations.
     Needless to say, these novel vacuum structures lead to many new potential
    applications for supersymmetric field theories, ranging from
   the cosmological-constant problem to the string landscape, 
    supersymmetry breaking, and $Z'$ phenomenology. 
\end{abstract}

\maketitle

\section{Introduction}

The vacuum structure of any physical theory plays a significant 
and often crucial role in determining the physical properties of that theory. 
In this talk, we describe three hitherto-unexplored and 
highly non-trivial vacuum structures which can arise in supersymmetric field
theories.

The first scenario we present~\cite{Nest} is an example of a metastable supersymmetry-breaking 
model in which all relevant features 
arise at tree-level in a completely calculable, perturbative 
framework.  
These include
a supersymmetric, $R$-symmetry-preserving 
ground state;  a metastable state in which both supersymmetry and $R$-symmetry are broken; 
and a vacuum energy barrier between the two of a sort that results in a long lifetime 
for the metastable vacuum. 
Neither the ground state nor the metastable vacuum involve runaways or 
flat directions;  moreover, 
the salient features of
the vacuum potential are perturbative and robust against quantum corrections, and the 
lifetimes of metastable vacua can be calculated reliably.
As far as we are aware, the model we present is the first such model with these
properties presented in the literature.

The second 
vacuum structure~\cite{GreatWalls} 
we present consists of large (and even infinite)
towers of metastable vacua, each with a distinct vacuum energy and
particle spectrum.  
We present a series of models which realize this vacuum structure explicitly.
As the number of vacua grows towards infinity in such models,
the energy of the highest vacuum stays fixed while the energy of the ground
state tends towards zero.
The instanton-induced tunneling dynamics associated with such vacuum towers results 
in a variety of distinct decay patterns;  these include not only 
regions of vacua experiencing direct collapses and/or tumbling cascades,
but also other regions of vacua whose stability is protected by ``great walls'' 
as well as regions of vacua populating ``forbidden cities'' into which tunnelling cannot occur. 

Finally,
the third vacuum structure we discuss~\cite{GreatWalls,toappear} arises as a limiting case of the
previous scenario.  In this limit, all of the metastable vacua in a given vacuum tower become degenerate,
and a shift symmetry emerges relating one vacuum to the next.
In such a scenario,
the true ground states of such theories are therefore nothing but Bloch waves
across these degenerate ground states, with energy eigenvalues approximating 
a continuum and giving rise to a vacuum ``band'' structure.  

Needless to say, 
these novel vacuum structures give rise to many new potential
applications for supersymmetric field theories, ranging from
the cosmological-constant problem to the string landscape, supersymmetry breaking, and $Z'$ phenomenology.
In this talk, we shall merely present these three different vacuum structures and
briefly sketch some possible applications;
further details can be found in Refs.~\cite{Nest,GreatWalls,toappear}.

\section{I.~~  Tree-Level Metastable Supersymmetry-Breaking}

The first model we discuss is   
a simple one
which can be used as a ``kernel'' for developing more complete models of 
metastable supersymmetry-breaking.  This model
consists of 
two $U(1)$ gauge groups, $U(1)_1$ and $U(1)_2$ with couplings $g_1=g_2=g$, 
and five chiral superfields $\Phi_i$, 
$i=1,...,5$, with charge assignments
$(-1,0)$, 
$( 1,-1)$, 
$(0,1)$, 
$(1,1)$, 
and $(-1,-1)$ respectively.
Given these charges, the most general renormalizable 
superpotential is given by 
\begin{equation}
             W~=~\lambda\,\Phi_1\Phi_2\Phi_3 \, +\, m\, \Phi_4\Phi_5~.
\label{eq:Wnest}
\end{equation}
We shall also generally permit non-zero Fayet-Iliopoulos terms $\xi_a$ for
each $U(1)_a$ gauge group.
Note that the $m=0$ version of this model was originally
discussed in Ref.~\cite{DDG} 
as part of a study of the string-theory landscape.

This model has a rich vacuum structure depending on the particular choices for the
parameters $(\lambda,m,\xi_a,\xi_b,g)$.
In general, the 
scalar potential of such a theory includes both $F$-term and $D$-term 
contributions and can be written in the form
\begin{equation}
      V~=~ \half \sum_a g_a^2 D_a^2 + \sum_i  |F_i|^2~
\label{eq:Vgen}
\end{equation}
where $D_a = \xi_a+\sum_i Q_{ai}|\phi_i|^2$ 
(with $\phi_i$ the scalar component of $\Phi_i$),
where $F_i \equiv -\partial W^{\ast}/\partial\phi^{\ast}_i$,
and where
$Q_{ai}$ is the $U(1)_a$-charge of $\Phi_i$. 
For concreteness here, however, let us
focus on the special case
$(\lambda,m,\xi_a,\xi_b,g) =(1,1,5,0,1)$.
We then find that
the scalar potential $V$ has three relevant critical points:  
two stable minima and one saddle point between them.
These occur at the specific {\it finite}\/ field-space locations 
$v_i\equiv \langle \Phi_i\rangle$ 
shown in Table~\ref{tab:solsprops}.

\begin{figure}[b!]
  \includegraphics[height=.33\textheight]{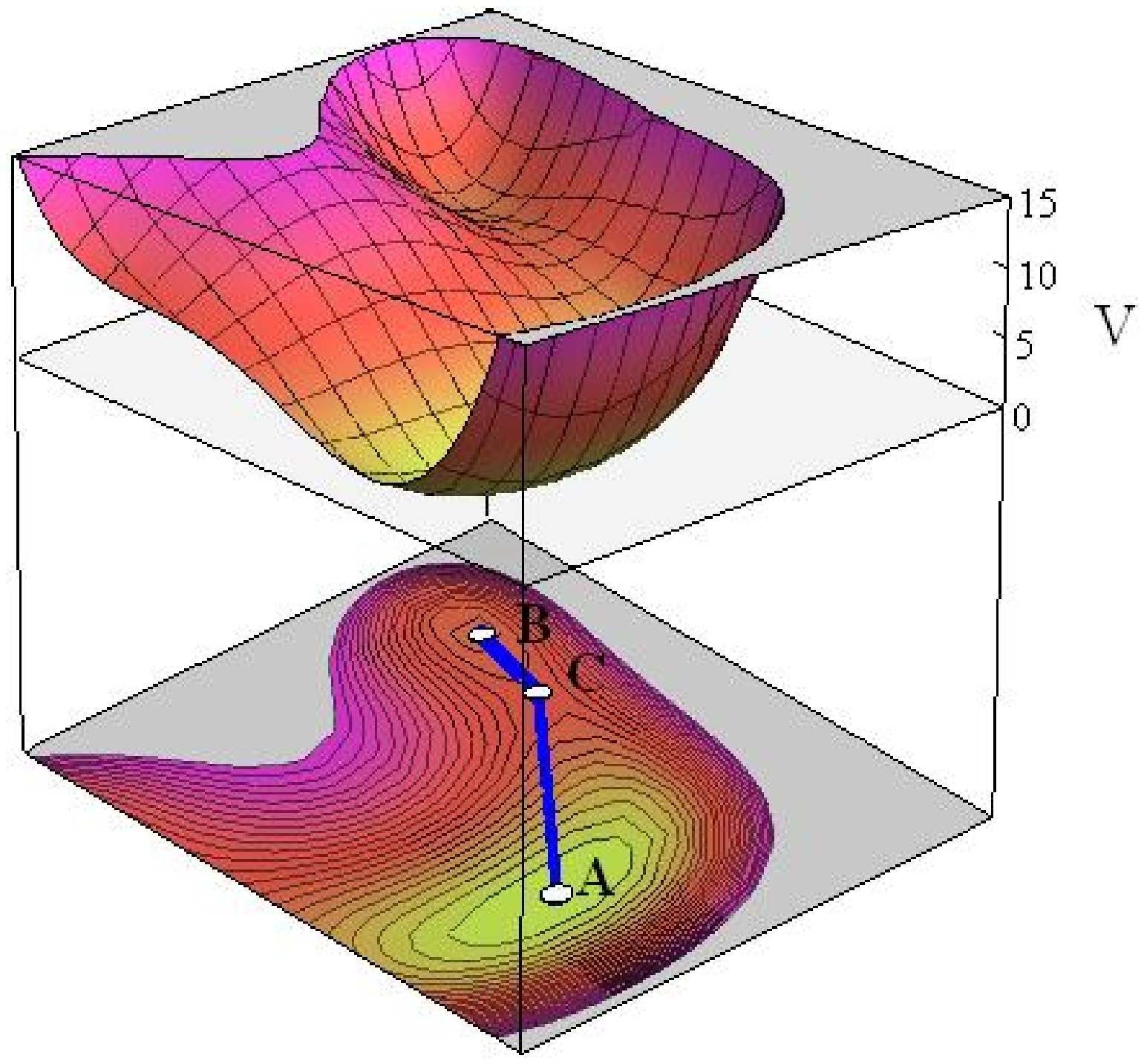}
  \includegraphics[height=.33\textheight]{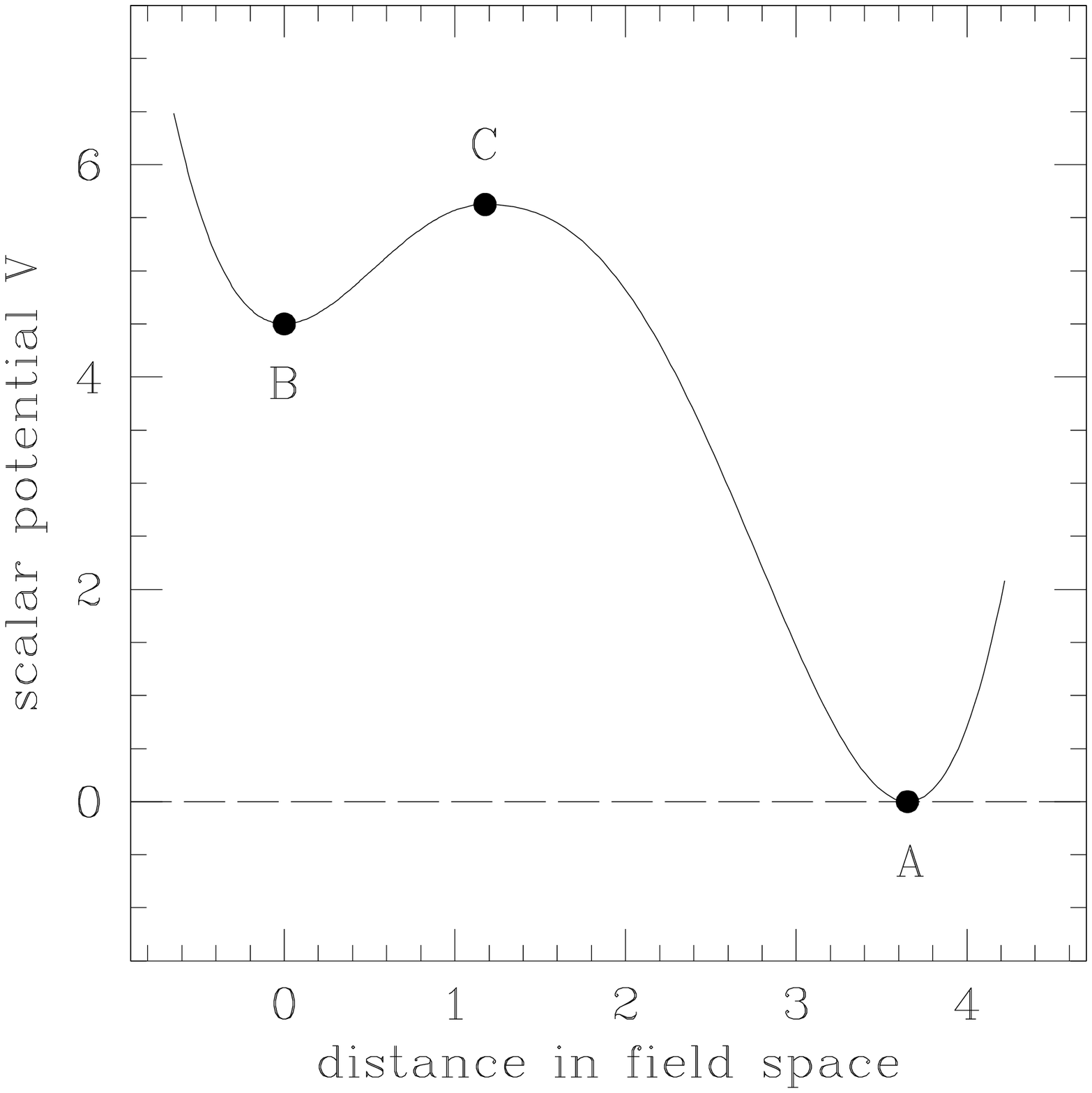}
  \caption{
           {\it Left figure}\/:  
      A surface plot of the scalar potential $V$ evaluated on
       the unique two-dimensional plane within the three-dimensional
      $(v_1,v_3,v_5)$ field space which simultaneously contains 
      the true vacuum solution~A
     the metastable vacuum solution~B, 
          and the saddle-point solution~C between them.
      Projected below the surface plot is a contour plot for $V$,
      showing the shortest path (blue) in field space connecting these three
      solutions.
      {\it Right figure}\/:
       The scalar potential $V$ evaluated along this shortest path.
     Field-space distances are quoted relative to the 
        metastable vacuum~B along this path in dimensionless units.
\label{fig:ABC}} 
\end{figure}

As we see from Table~\ref{tab:solsprops},
Solution~A represents a supersymmetric,
$R$-symmetry-preserving ground state with vanishing vacuum energy, while Solution~B
represents a metastable minimum in which supersymmetry and $R$-symmetry are both broken.
Solution~C represents the saddle-point solution through which the classical path between
the two vacua passes.  This vacuum structure is shown explicitly in Fig.~\ref{fig:ABC}.    
  
\begin{table}[t!]
\begin{tabular}{ccccccc}
\hline
    \tablehead{1}{r}{b}{Label}
  & \tablehead{1}{c}{b}{$(v_1,v_3,v_5)$}
  & \tablehead{1}{c}{b}{$V$}
  & \tablehead{1}{c}{b}{Stability}
  & \tablehead{1}{c}{b}{SUSY}
  & \tablehead{1}{c}{b}{$R$-symmetry}
  & \tablehead{1}{c}{b}{Gauge Group}   \\
\hline
     A & $(\sqrt{5},0,0)$               & $0$       & Stable     & Yes & Yes & $U(1)_b$  \\
     B & $(0,2,2)$                      & $9/2$     & Metastable & No  & No  & None  \\
     C & $(\sqrt{3}/2, \sqrt{7}/2,\sqrt{5/2})$ & $45/8$    & Unstable   & No  & No  & None  \\ 
    \hline
\end{tabular}
\caption{The classical vacuum structure of the model in Eq.~(\ref{eq:Wnest})     
       with $(\lambda,m,\xi_a,\xi_b,g)=(1,1,5,0,1)$. 
       Here Solutions~$A$ and $B$ correspond to stable vacua, while $C$ corresponds
       to a saddle point.  For each of these solutions, we have listed 
       the corresponding field VEVs $(v_1,v_3,v_5)$, along with 
       the value of the scalar potential,
       the stability and supersymmetry properties of that extremum,  
       and the surviving (unbroken) gauge group.
       Note that each solution has $v_2=v_4=0$.
       We have also indicated whether
       $R$-symmetry is broken or unbroken at that extremum,
       assuming that each of the chiral superfields in this model carries
       an appropriate non-zero $R$-charge discussed in Ref.~\protect\cite{Nest};
       likewise, that all dimensionful quantities are
       quoted in dimensionless units.}
\label{tab:solsprops}
 \end{table}

In order to be of phenomenological interest for model-building, the lifetime 
of any given metastable vacuum must be at least on the order of the present 
age of the universe.  This lifetime can be determined using standard instanton 
methods~\cite{Instantons,Triangles}, and it can be shown that 
the metastable minimum in our model is stable on cosmological time scales
over a large region of parameter
space, including the point $(\lambda,m,\xi_a,\xi_b,g)=(1,1,5,0,1)$ discussed above.
These calculations are discussed more fully in Ref.~\cite{Nest}.

As we have shown,
the supersymmetry and $R$-symmetry in our construction are broken at tree level in a
perturbative theory where no flat directions appear in
the classical potential and where all minima appear at finite distances
in field space.
Thus, our construction can be viewed as an alternative 
to those which have appeared in much of the prior
literature on metastable supersymmetry-breaking (for original papers, 
see, \eg, Ref.~\cite{originals}).
Indeed, most previous models actually give rise to ``vacua''
containing
either classical flat
directions or runaway behavior.
Compared with previous constructions,
the key feature of our model is that the supersymmetry-breaking
arises not only through $F$-term breaking, but also through $D$-term breaking.
Since all of the relevant physics is perturbative,
we are able to perform
explicit calculations of the lifetimes and particle spectra associated with such vacua
and demonstrate that these lifetimes can easily
exceed the present age of the universe.
Models which have these properties are extremely rare,
and we are not aware of any models with these
characteristics in the prior metastability literature.

\section{II.~~  Metastable Vacuum Towers}

The second scenario~\cite{GreatWalls} 
we discuss involves a class of models which give rise to towers   
of non-degenerate, metastable vacua. 
The underlying structure of these models
is that of 
an $N$-site orbifold Abelian moose consisting of $N$ different $U(1)$ gauge 
groups with a common coupling $g$ and $N+1$ chiral superfields $\Phi_i$.   
To this structure, we then add three critical ingredients, each of which
is vital for the emergence of our metastable vacuum towers.  The first of these
is the introduction of a single Wilson-line operator
\begin{equation}
            W~=~\lambda\, \prod_{i=1}^{N+1}\Phi_i~
\label{eq:Wtower}
\end{equation}
which represents the most general superpotential that can be formed from the fields of the
theory.  The second and third ingredients both exploit the Abelian nature of our 
gauge groups:
one is to introduce non-zero Fayet-Iliopoulos terms $\xi_1$ and $\xi_N$ for the 
``endpoint'' gauge groups $U(1)_1$ and $U(1)_N$ respectively, while the other is to introduce
kinetic mixing~\cite{Holdom} among the various $U(1)$ factors in the theory.  
The effect of including 
kinetic mixing is that the gauge-kinetic part of the Lagrangian is modified to include a
mixing matrix $X_{ab}$ and hence takes the form~\cite{SUSYKineticMixing} 
\begin{equation}
    {\cal L}~\ni~\frac{1}{32}\int d^2\theta ~~ W_{a\alpha} \, X_{ab} \, W_b^{\alpha}~
\label{eq:KMixLagrange}
~~~~~{\rm where}~~		
       X_{ab}\equiv \pmatrix{
             1 & -\chi_{12} & \ldots & -\chi_{1N} \cr
                 -\chi_{12} & 1 & \ldots & -\chi_{2N} \cr
                 \vdots & \vdots & \ddots & \vdots \cr
                 -\chi_{1N} & -\chi_{2N} &  \ldots & 1 \cr}~.
\end{equation} 
To simplify our analysis, we will focus on the case in which mixing occurs 
only between nearest-neighbor sites on the moose, with a common mixing 
parameter $\chi$ ({\it i.e.}\/, $\chi_{ab}=\chi\delta_{a,b+1}$);   likewise, we shall
assume that 
$0<\chi<1/2$.
We shall also take
$\xi_1=\xi_N\equiv \xi$  for simplicity.
Note that a similar model, but with $\chi=0$ and $\lambda\to0$, was previously considered
in Ref.~\cite{DudasDecon,DDG}.

\begin{figure}
  \includegraphics[height=.4\textheight]{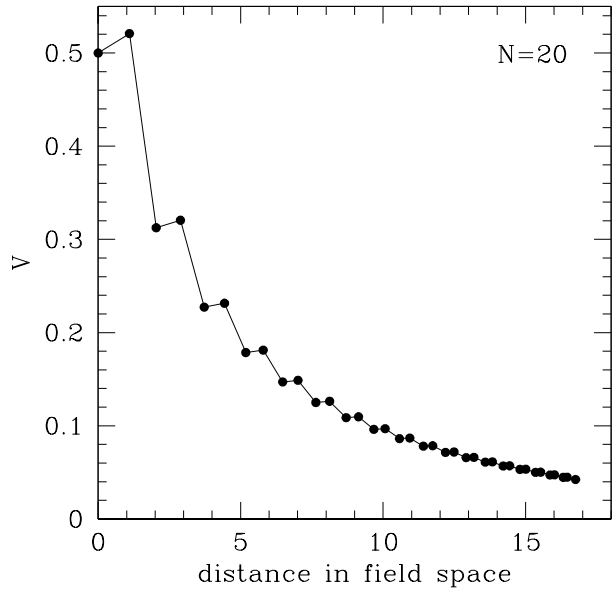}
  \caption{The vacuum structure of the $N=20$ model, plotted for $\chi=1/5$.
      This model gives rise to a tower of 18 metastable vacua above the
      true ground state.
      Vacuum energy is plotted on the vertical axis,
      while the horizontal axis indicates the cumulative distances in field space
      along a trajectory which begins at the $n=1$ vacuum and then proceeds
      along straight-line path segments to the $(1,2)$ saddle point, then to
      the $n=2$ vacuum, then to the $(2,3)$ saddle point, and so forth.} 
\label{fig:N20plots}
\end{figure}

Given this setup,
we find that 
the vacuum structure of this model contains $N-1$ stable
vacua.  These can be labelled with an index $n=1,\cdots,N-1$ in order of
decreasing vacuum energy.  We then find that the vacuum energy of the 
$n$-vacuum is given by
\begin{equation}
         V_n ~=~   \half \left( 1\over \chi R_n\right)~~~~~~~~~{\rm where}~~
            R_n~\equiv~ \left({1\over \chi} - 2\right)n + 2~,
\label{Vn}
\end{equation}
and corresponds to the solution with
\begin{eqnarray}
v_j^2 = \cases{
         1+ 1/R_n  &  for $j=1$\cr
    1/R_n        & for $2\leq j \leq N-n$\cr
    0            & for $j=N-n+1$\cr
   (R_{j-N+n-1}-1)/R_n~~~ & for $N-n+2\leq j \leq N$ \cr
    0             &  for $j=N+1$~ \cr}
\label{vacsolns}
\end{eqnarray}
where we continue to list all quantities in dimensionless units and
where all dependence on the overall coupling $g$ has been rescaled away.
This vacuum tower is illustrated for the $N=20$ case in Fig.~\ref{fig:N20plots}.

As required, these vacua are separated from one another by 
saddle-point solutions in field space.  
The general expressions for the
field-space coordinates and barrier heights associated with these saddle points 
are generally quite complicated,
but they are controlled primarily 
by the Wilson-line coefficient $\lambda$.
As a result,
the stability of any given vacuum state in the tower is therefore also primarily 
governed by $\lambda$:   indeed, for any $N$, we find that 
the $n$-vacuum
will be stable as long as $\lambda$      
exceeds the critical value
\begin{equation}
    \lambda_{N,n}^{\ast 2} ~=~   
           y^n ~ {\Gamma(y)\over \Gamma(n+y)}
                ~{R_n^{N-2} \over \chi (1+R_n)}~.
\label{lambdastarNn}
\end{equation}
Here $y\equiv \chi/(1-2\chi)=n/(R_n-2)$ and 
$\Gamma(z)$ is the Euler $\Gamma$-function [for which
$\Gamma(z)= (z-1)!$ when $z\in\IZ^+$]. 
Consequently, 
all of the vacua in the tower will be stable
as long as $\lambda$ exceeds the maximum value of $\lambda_{N,n}^\ast$.
Moreover, there is nothing which prevents us from taking $N\to\infty$ in all of our results.
As a result, we see that we can achieve a vacuum structure containing
literally an infinite tower of metastable vacua. 

\begin{figure}
  \includegraphics[height=.25\textheight]{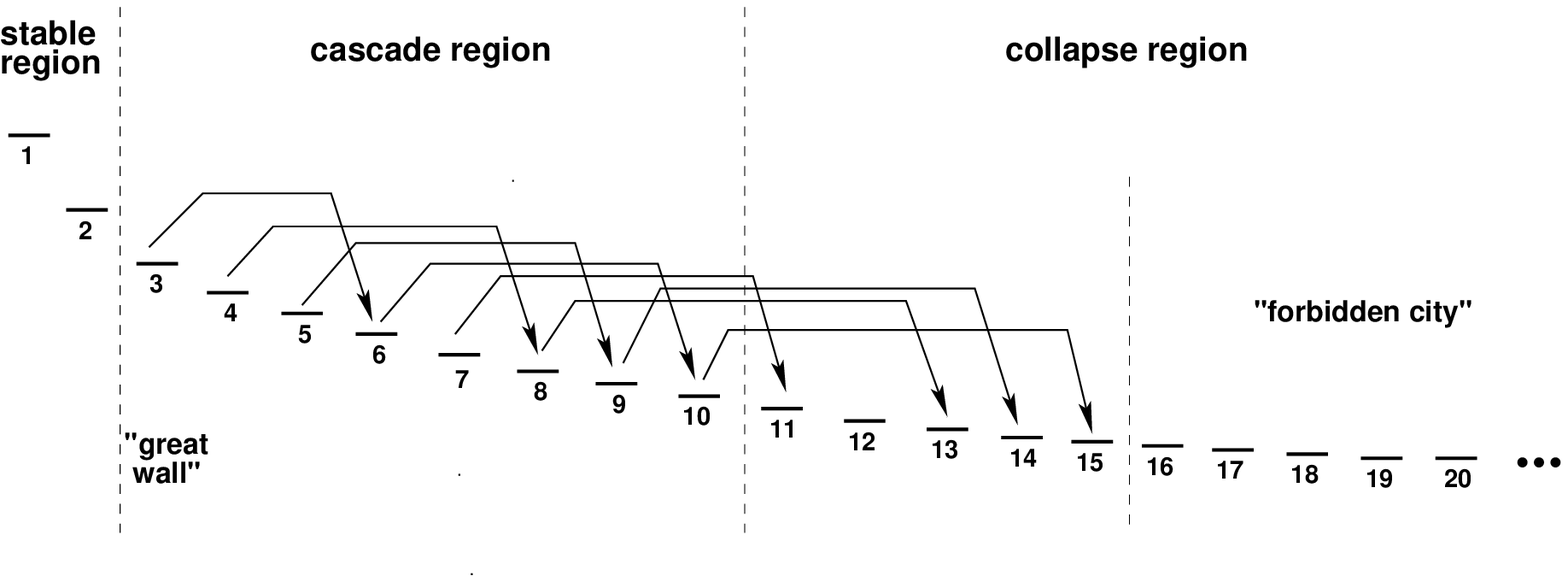}
  \caption{A schematic of the vacuum
    cascade dynamics that arises in a model with $N=5000$ and $\chi=2.8\times 10^{-4}$.
    Vacua in the stable region to the left of the ``Great Wall'' have lifetimes 
    exceeding the age of the universe, while vacua in the cascade region
    decay to other (lower) metastable vacua in the vacuum tower.  By contrast, vacua in the 
    collapse region decay directly to the ground state
    of the vacuum tower.  Finally, vacua which populate
    the ``Forbidden City'' cannot be reached from outside the Forbidden City:  such vacua
    can be populated only as an initial condition at the birth of the vacuum configuration.}
\label{fig:cascade}
\end{figure}

Given the existence of such vacuum towers, 
we find that instanton-induced tunneling can produce a
variety of highly non-trivial vacuum decay patterns.
These include not only 
regions of vacua experiencing direct ``collapses'' 
(in which a given metastable vacuum decays directly to the ground state)
and/or tumbling ``cascades''
(in which a given metastable vacuum decays to another metastable vacuum, and so forth).
There are also other regions of vacua whose stability is protected by ``great walls''
as well as regions of vacua populating ``forbidden cities'' into which tunnelling cannot occur.

One example~\cite{GreatWalls} which illustrates all of these 
features 
simultaneously is shown in Fig.~\ref{fig:cascade}.
In this example,
the $n=3$ vacuum decays into the $n=6$
vacuum, which in turn
decays (even more rapidly) into the $n=10$ vacuum;
this in turn decays (even more rapidly) into the $n=15$ vacuum,
which in turn decays directly into the ground state.
In fact, in this specific example, there are  
four independent potential cascade trajectories,
each of which unfolds with increasing speed (\ie, decreasing lifetimes):
\beqn
   && \bullet ~~3 \to 6 \to 10 \to 15 \to {\rm GS} \nonumber\\
   && \bullet ~~4 \to 8 \to 13 \to {\rm GS}\nonumber\\
   && \bullet ~~5 \to 9 \to 14 \to {\rm GS}\nonumber\\
   && \bullet ~~7 \to 11 \to {\rm GS}
\eeqn
where `GS' signifies the ground state.
It is therefore only an initial condition that determines which trajectory a
given system ultimately follows.

There are, of course, limits to this cascade region, both at the top
and at the bottom.
For example, the top two vacua
have decay rates which fall below those needed for cosmological stability;
these vacua,
if initially populated, are therefore deemed stable on cosmological time scales.
Likewise, at the $n=11$ vacuum and beyond, we enter the collapse region
in which all subsequent decays automatically proceed directly to the ground state.

Clearly, there are also a number of possible applications for a vacuum
structure of this sort.
Perhaps the application which most immediately springs to mind concerns
a potential solution to the cosmological-constant problem.
Over the past decade, several scenarios have been proposed
in which a small cosmological constant emerges as a consequence of a large
number of vacua~\cite{BoussoPolchinski,banks,Gordy,tye}.  Scenarios of this sort
tend to posit the existence of a ``landscape'' of vacua with certain gross
properties, including a vacuum state whose energy is nearly vanishing.
One then imagines that the universe either dynamically tumbles down
to this special vacuum state, or is somehow born there.
However, to the best of our knowledge, no explicit model with a vacuum
structure exhibiting such properties has ever been constructed.
It is our hope that the model we have presented here
might provide an explicit field-theoretic realization of such a scenario.

Another potential implication of such a vacuum structure concerns
statistical studies~\cite{douglas} of the string landscape~\cite{Susskind}.
One important issue that needs to be addressed when performing such studies
concerns the proper definition of a {\it measure}\/
across the landscape:  in what manner are the different string
theories to be weighted relative to each other?
Clearly, the most na\"\i ve approach is to count each string
model equally, interpreting each as contributing a single vacuum
state to the landscape as a whole.  However, 
moose theories of the sort we have been discussing here
often appear as the actual low-energy (deconstructed) limits of
flux compactifications~\cite{DDG}, and as we have seen, such
theories give rise to infinite towers of metastable vacua.
Thus, if the true underlying landscape measure is based
on {\it vacua}\/ rather than {\it models}\/, then a theory with 
infinite towers of vacua is likely
to dominate any statistical study of the string landscape.
As such, the phenomenological properties of these sorts of models
will dominate the properties of the landscape as a whole.

Further potential applications of such scenarios involve
supersymmetry breaking, $Z'$ phenomenology,
and cosmological evolution.
Full discussions of these and other ideas can be found
in Ref.~\cite{GreatWalls}.

\section{III.~~  Degenerate Vacua and Bloch Waves}

The results presented in Sect.~II concerning our $N$-site moose 
are applicable in the range $0< \chi< 1/2$, where $\chi$ is our kinetic-mixing parameter.
Indeed, each successive vacuum in the resulting vacuum tower has a lower energy than
the previous one, and consequently there exists a net direction
for dynamical flow.

\begin{figure}[b!]
  \includegraphics[height=.4\textheight]{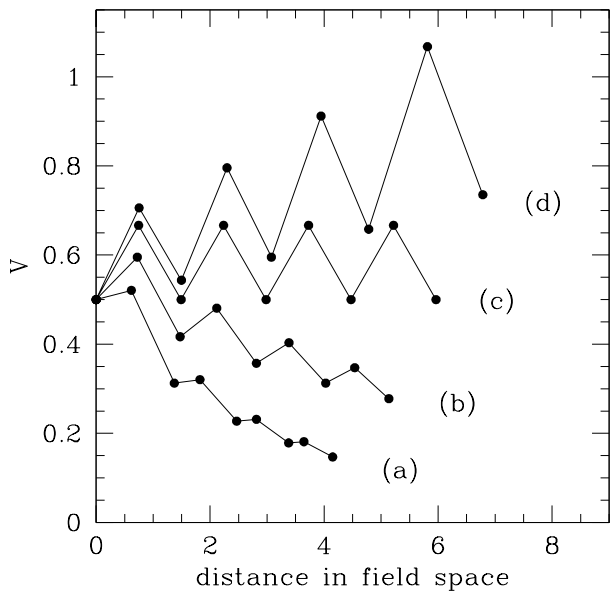}
  \caption{The manner in which the vacuum structure of our model depends on $\chi$.
Here we have focused on the $N=6$ model, 
and taken (a) $\chi=0.2$, (b) $\chi=0.4$, 
(c) $\chi=0.5$, and (d) $\chi=0.54$.
As $\chi$ increases from zero, we see that
the ``slope'' of our vacuum ``staircase''
decreases, ultimately becoming completely flat 
at $\chi=0.5$.
\label{fig:ShowBloch}}
\end{figure}

However, in the $\chi\to 1/2$ special case, a new behavior develops.
For $\chi=1/2$, 
we find that $R_n=2$ for all $n$, and thus the expressions for the vacuum 
energies $V_n$ 
become independent of the vacuum index $n$.  As a result, the
resulting vacuum structure consists of $N-1$ {\it degenerate}\/ 
vacua.  These
turn out to be  separated 
from each other by a set of equivalent saddle-point potential barriers of 
uniform height. 
The field-space distances between all pairings of vacua also
become equal in this case.
This behavior is shown in Fig.~\ref{fig:ShowBloch}.

As a result,
the true ground state of such a theory when $\chi=1/2$ is no
longer any of the individual $n$-vacua by itself.  Instead, what emerges is an
infinite set of {\it Bloch waves}\/ across the entire set of degenerate vacua.
Moreover, the vacuum energies associated with such Bloch waves fill out a continuous
band.  The 
vacuum energy of the true ground state of the theory
will consequently be smaller than the vacuum energy of any individual vacuum.

It is interesting to speculate that the vacuum energy of this
true ground state might actually vanish (thereby {\it restoring}\/ supersymmetry) 
or alternatively merely {\it approach}\/ 
zero as $N\to\infty$ (yielding a a very small cosmological constant
in a manner reminiscent of the proposal in Ref.~\cite{Gordy}).
It is also interesting to note that regardless of the ground-state energy,
the fact that our Bloch vacuum states are linear combinations of our individual $n$-vacua
implies that a system originally populating a given $n$-vacuum
will experience non-trivial {\it time-dependent oscillations}\/ across  
the set of $n$-vacua as a whole.
Such vacuum oscillations are analogous to multi-flavor neutrino oscillations,
and can potentially have dramatic effects on the phenomenology resulting
from such theories.
These and other issues will be explored more fully in Ref.~\cite{toappear}.

\begin{theacknowledgments}
We would like to thank 
the organizers of the XIII Mexican School of Particles and Fields,
especially
Alejandro Ayala and 
Maria Elena Tejeda-Yeomans, 
for their warm hospitality and invitation to attend and present this work.
Moreover, we were happy to discover that the Sea of Cortez --- 
just like the Great Wall and Forbidden City of China ---
is capable of providing 
abundant inspiration to desert-dwelling creatures like ourselves.
This work was 
supported in part by the Department of Energy under Grant~DE-FG02-04ER-41298,
and no red wheelbarrows were harmed during the delivery of this talk.
\end{theacknowledgments}

\bibliographystyle{aipprocl} 

\end{document}